\documentclass{aastex}
\usepackage{emulateapj5,apjfonts}

\newcommand{\etal}{et al.{}}
\newcommand{\LCDM}{{\mbox{$\Lambda$CDM}}}
\newcommand{\Msun}{{\mbox{$M_\odot$}}}

\shorttitle{Radiative cooling and X-ray clusters}
\shortauthors{Muanwong \etal}

\begin{document}

\title{The effect of radiative cooling on scaling laws of X-ray 
groups and clusters}

\author{O. Muanwong\altaffilmark{1},
        P. A. Thomas\altaffilmark{1}, 
        S. T. Kay\altaffilmark{1},
        F. R. Pearce\altaffilmark{2} and 
        H. M. P. Couchman\altaffilmark{3}
        }
\altaffiltext{1}{Astronomy Centre, University of Sussex, Falmer, Brighton,
  BN1\,9QJ, U.K.; O.Muanwong@sussex.ac.uk, P.A.Thomas@sussex.ac.uk,
  S.T.Kay@sussex.ac.uk}
\altaffiltext{2}{Dept.~of Physics, University of Durham, South Road, Durham,
  DH1\,3LE, U.K.; F.R.Pearce@durham.ac.uk}
\altaffiltext{3}{Dept.~of Physics, McMaster University, Hamilton, Ontario,
  L8S\,4M1, Canada; couchman@physics.mcmaster.ca}

\begin{abstract}
We have performed cosmological simulations in a $\Lambda$CDM cosmology
with and without radiative cooling, in order to study the effect of
cooling on the cluster scaling laws. Our simulations consist of
4.1 million particles each of gas and dark matter within a box-size of
100 $h^{-1}$\,Mpc and the run with cooling is the largest of its kind
to have been evolved to $z=0$. Our cluster catalogues both consist of
over 400 objects and are complete in mass down to $\sim 10^{13} h^{-1}
{\rm M_{\odot}}$.  We contrast the emission-weighted temperature-mass
($T_{\rm ew}-M$) and bolometric luminosity-temperature ($L_{\rm
bol}-T_{\rm ew}$) relations for the simulations at $z=0$. We find that
radiative cooling increases the temperature of intracluster gas
and decreases its total luminosity, in agreement with the
results of Pearce et al. Furthermore, the temperature dependence of
these effects flattens the slope of the $T_{\rm ew}-M$ relation and
steepens the slope of the $L_{\rm bol}-T_{\rm ew}$ relation.
Inclusion of radiative cooling in the simulations is sufficient to reproduce
the observed X-ray scaling relations without requiring excessive
non-gravitational energy injection.
\end{abstract}


\section{Introduction}
\label{sec:introduction}

The mass of clusters of galaxies\footnote{Throughout this paper we
will not distinguish between groups and clusters of galaxies but will
use the term clusters to stand for both} is dominated by dark matter.
The evolution of the dark matter halo population is now
well-understood both theoretically \citep{LaC94} and numerically
\citep{JFW01}.  The halos themselves are approximately self-similar
and may be described in their inner regions by a one-parameter model
\citep{NFW97} with a concentration parameter that is a slow function
of mass. (Note, however, that there are significant deviations from this
simple profile in the outer parts of clusters; see Thomas et~al.\
2001.\nocite{TMP01})

The intracluster medium (ICM) does not share the approximate
self-similarity of the dark matter.  This is expressed most clearly in
the X-ray luminosity-temperature relation.  For pure bremsstrahlung
emission, the bolometric X-ray luminosity should scale with
temperature as $L_{\rm x}\propto T_{\rm x}^2$ (the inclusion of line emission
flattens this relation), whereas observations indicate a much steeper
temperature dependence, especially for low-mass systems \citep{EdS91,
DSJ93,PBE96,WJF97,XuW00}.  At high temperatures part of this
steepening is due to a central cooling flow, but removing the cooling
flow component still does not reconcile the observations with the
self-similar prediction \citep{AlF98,Mar98}.

The reason for the departure from self-similarity is that the gas is
not as centrally-concentrated in clusters as the dark matter, which is
best physically expressed as an increase in entropy of the innermost
gas \citep{EvH91,Kai91,Bow97}. 
The most obvious explanation for this is that there has been some form
of energy injection.  The amount of energy required depends upon the
density of the gas at the time when the heating occured.
\citet{LPC00} argue for heating prior to cluster collapse and
estimate a value of 0.3\,keV per particle.  \citet{ToN01} show
that this can be lowered to 0.1\,keV per particle by heating at the
optimal time (when the gas is at its minimum density) but most other
studies that consider heating within collapsed halos require much
higher values of 1--3\,keV per particle \citep{WFN99,BBB00,Low00}.

There are two likely sources for any excess energy: stellar
winds/supernovae and active galactic nuclei (AGN).  It is known for
CDM models that some feedback of energy into the intergalactic medium
must occur in order to prevent the ``cooling catastrophe'', in which
the majority of the baryons in the Universe cool and form stars at
high redshift \citep{WhF91,Col91,BVM92}.  If the heating efficiency is
high, supernovae can inject an energy of order 0.3\,keV per particle
into the intergalactic medium, but they do not do so in an optimal
way.  Various authors, all of whom consider realistic, but different,
models for the build-up of structure, conclude that supernovae are
unable to provide the required excess entropy
\citep{VaS99,WFN99,BBB00}.  Heating of the intergalactic medium by
quasars is also not without its problems as the heating must occur at
just the right time in order not to overly suppress galaxy formation.
Alternatively, the heating may arise from AGN buried within individual
galaxies \citep{BBB00}.

Radiative cooling results in the removal of low-entropy gas in the
cluster core and thus leads to an overall increase in temperature of
the ICM \citep{ThC92}.  The amount of cooling that takes place in a
cooling flow after the final assembly of a cluster is insufficient to
explain the observations \citep{BBB00}, but a more realistic model in
which cooling occurs at every stage of the collapse hierarchy can give
entropy increases equivalent to an excess energy of 1--2\,keV per
particle \citep{WFN99}---in essence most of the cooling occurs in
galaxy-sized halos before the formation of the cluster. \citet{Bry00}
has developed a simple model in which low-entropy gas is removed from
the cluster core and the surrounding gas is assumed not to have cooled
at all.  Although this model has obvious deficiencies, it predicts the
correct luminosity-temperature relation from 0.5--10\,keV.

Simulations of cluster formation including radiative energy loss have
been carried out by \citet{PTC00}.  They showed that the gas is
slightly heated and that the luminosity is greatly reduced (except in
cooling flow clusters), in agreement with expectations.  However, the
simulations were of limited resolution and covered only a small range
in cluster mass.  We are now undertaking a programme of simulations to
extend these results over a wider mass-range and to contrast the
properties of clusters in different cosmologies.  In this letter we
report results at $z=0$ from two simulations of a 100\,$h^{-1}$Mpc box
in the \LCDM\ cosmology, one with and one without radiative cooling.

The simulations and cluster extraction method are described in
Section~\ref{sec:method} and the results are presented in
Section~\ref{sec:results}.  We summarize our conclusions in 
Section~\ref{sec:conclusions}.

\section{method}
\label{sec:method}

\subsection{The simulations}
\label{sec:simulations}

We have carried out two simulations with 160$^3$ particles each of gas
and dark matter within cubical volumes of side 100\,$h^{-1}$Mpc.  The
cosmological parameters were as follows: density parameter,
$\Omega_0=0.35$; cosmological constant,
$\Lambda_0=\Lambda/3H_0^2=0.65$; Hubble parameter,
$h=H_0/100$km\,s$^{-1}$Mpc$^{-1}=0.71$; baryon density parameter,
$\Omega_{\rm b}h^2=0.019$; power spectrum shape parameter, $\Gamma=0.21$;
and a linearly-extrapolated root-mean-square dispersion of the density
fluctuations on a scale 8\,${h^{-1}}$Mpc, ${\sigma_8}= 0.90$.  With
these parameters, the gas and dark matter particle masses are approximately
$2.6\times10^9$ and $2.1\times10^{10}h^{-1}{\rm M_{\odot}}$,
respectively.  This mass is below the \citet{StW97} limit above
which numerical heating dominates cooling.
The runs were started at a redshift, $z=50$ and evolved to the present day, 
$z=0$.   The gravitational softening was fixed at $50\,h^{-1}$kpc in comoving 
co-ordinates until $z=1$ and thereafter held constant at 
25$\,h^{-1}$kpc in physical co-ordinates.  This softening is
sufficient to prevent two-body relaxation \citep{TMP01} and we have
checked that the gas and dark-matter have similar specific energy
profiles in clusters drawn from the non-radiative simulation.

The only difference in the two runs was that one of them included
radiative cooling.  For this run, we used the cooling tables of
\citet{SuD93} and assumed a uniform but time-varying metallicity of
$Z=0.3\,(t/t_0)\,Z_\odot$, where $t/t_0$ is the age of the Universe in
units of the current time.  This time-varying metallicity is meant to
crudely mimic the gradual enrichment of the ICM by stars, but more
importantly it results in a global cooled gas fraction at the end of
the simulation of approximately 20 per cent.  This is the maximum
value inferred from the observations of clusters \citep{BPB01} which
means that our simulation will represent the largest effect that
cooling alone is likely to have on the intracluster medium.

We use a parallel version of the {\sc hydra} $N$-body/SPH code as
described by \citet{CTP95} and \citet{PeC97} except that the SPH
equations have been modified to use the pairwise artificial viscosity
of \citet{MoG83}---for a test of different SPH formalisms see
\citet{TTP00}.  We decouple the hot and cold gas in the manner
described by \citet[see also Ritchie \& Thomas 2001
\nocite{RiT01}]{PJF99} to prevent artificial overcooling of hot gas
onto the central cluster galaxies.  Groups of 13 or more cold, dense
gas particles (with $\delta>500$ and $T<$12\,000\,K) are merged
together to form collisionless {\it galaxy} particles, which can only
grow via the accretion of more gas. Not only does this save
considerable computational effort, it also prevents small objects from
being artifically disrupted within cluster potentials.

\subsection{The cluster catalogue}
\label{sec:catalogue}

Initially we identify clusters in our simulation by searching for
groups of dark matter particles within an isodensity contour of 200,
as described in \citet{TCC98}.  We work with a preliminary catalogue
of all objects with more than 250 particles, then retain only those
which have a total mass within the virial radius exceeding $M_{\rm
lim}\approx1.18\times10^{13}h^{-1}\Msun$, corresponding to 500
particles of each species.  The use of a small mass for the
preliminary cluster selection ensures that our catalogue is complete.
We have checked that using a different isodensity threshold, a
different selection algorithm, or using gas particles instead of
dark-matter particles to define the cluster, leads to an almost
identical cluster catalogue---the only difference being the merger or
otherwise of a small number of binary clusters.

We define the centre of the cluster to be the position of the densest
dark matter particle. 
Because the density parameter of the real universe is not known, we
choose to average properties of the clusters within spheres that
enclose an overdensity of 200 relative to the critical density
(whereas for this cosmology the virial radius corresponds to an
overdensity relative to critical of about 110).  Our final catalogues
consist of 427 and 428 clusters in the radiative and non-radiative
simulations, respectively.

\subsection{Cluster X-ray properties}
\label{subsec:xrayprop}

We calculate the bolometric luminosity of each cluster using
\begin{equation}
L_{\rm bol} = \sum_{i} \, {m_i \, \rho_i \over (\mu m_{\rm H})^2} 
\Lambda(T_i,Z),
\label{eqn:lx}
\end{equation}
where the subscript $i$ denotes the sum over all gas particles within
radius $r_{200}$ that have temperatures, $T_i>10^{5}{\rm K}$, masses
$m_i$ and densities, $\rho_i$; we assume a mean molecular mass $\mu
m_{\rm H}=10^{-24}$g and an emissivity, $\Lambda(T_i,Z(t))$, that is
the same function used by {\sc hydra} to calculate cooling rates, as
discussed in Section~\ref{sec:simulations}.  Note that, for
temperatures below about $10^7$K, line emission is important and the
cooling rates are substantially higher than the contribution from
bremsstrahlung alone.

The emission-weighted temperature of each cluster is calculated as
\begin{equation}
T_{\rm ew} = {\sum_{i} \, m_i \, \rho_i \Lambda(T_i,Z) \, T_i
       \over \sum_{i} \, m_i \, \rho_i \Lambda(T_i,Z)}.
\end{equation}

Many of the smaller clusters in our catalogues have emission-weighted
temperatures that are below 0.5\,keV which means that the majority
of their emission will emerge at energies that are below the X-ray
bands.  We do not attempt to calculate the emission in any particular
X-ray passband in this paper but instead quote bolometric luminosities
and emission-weighted temperatures.

\section{Results}
\label{sec:results}

\subsection{Temperature-mass relations}
\label{sec:mt}

\vspace{\baselineskip} \epsfxsize=6.5cm
\centerline{\rotatebox{270}{\epsfbox{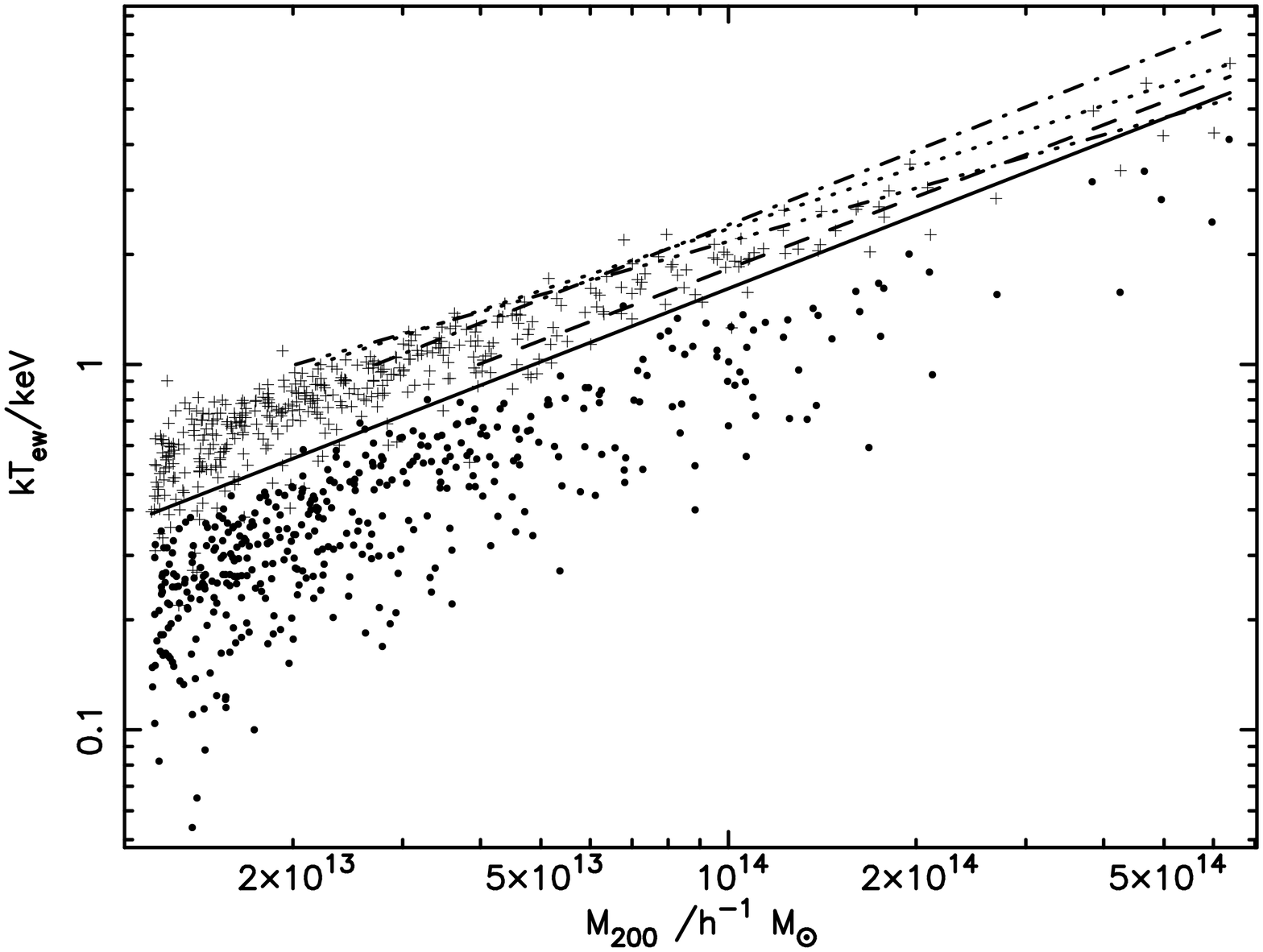}}} 
\figcaption{The emission-weighted temperature versus (total) mass
within a density contrast of 200.  Crosses and filled circles
represent clusters from the simulations with and without radiative
cooling, respectively.  The various lines are described in the text.
\label{fig:tm200}}
\vspace{\baselineskip}

We plot the emission-weighted temperature-mass relation for both
cluster samples in Fig.~\ref{fig:tm200}.  Results from the simulations
with and without cooling are illustrated using crosses and filled
circles, respectively. The broken lines are power-law fits to the
observational relation as determined by \citet{HMS99}, using mass
estimates from galaxy velocity dispersions (dashed), X-ray temperature
profiles (dash-dotted), the isothermal $\beta$-model (dotted) and the
surface brightness deprojection method (dash-triple dotted).

For the simulation with radiative cooling the relationship between ${T_{ew}}$
and $M$ is an approximate power law
\begin{equation}
\label{eq:mtxcool}
kT_{\rm ew} = 1.91\left(\frac{M_{200}}{10^{14} 
 h^{-1}{\rm M_{\odot}}}\right)^{0.58} {\rm keV},
\end{equation}
and for the simulation without radiative cooling 
\begin{equation}
\label{eq:mtxnocool}
kT_{\rm ew} = 0.98\left(\frac{M_{200}}{10^{14} 
 h^{-1}{\rm M_{\odot}}}\right)^{0.67} {\rm keV}.
\end{equation}
These should be compared with the virial relation (shown as
the solid line on the figure)
\begin{equation}
\label{eq:mtvir}
kT_{\rm vir} = 1.61\left(\frac{M_{200}}{10^{14} 
 h^{-1}{\rm M_{\odot}}}\right)^{0.67} {\rm keV}.
\end{equation}
(Note that no attempt has been made when making these fits to 
reproduce the observational selection effects. They should therefore
be regarded as rough guides rather than precise predictions.)
The clusters from the non-cooling run mostly have emission-weighted
temperatures that are much lower than the virial values.  The reason
for this is that their emission is dominated by high-density,
low-temperature gas in the core of the cluster (this temperature drop
in the core is simply due to the fact that the specific energy profile
of the gas mimics that of the dynamically-dominant dark matter).
Small fluctuations in the core properties of the clusters (which are
not well-resolved by our simulations) lead to a large scatter in
predicted temperatures.  It is important to note that these
non-radiative simulations do not provide sensible predictions for the
observed properties of real clusters---the core gas has a short
cooling time and would not in reality persist in the intracluster
medium for the lifetime of the cluster.

We note that our non-radiative $T_{\rm ew}-M$ relation has a lower
normalization than found by \citet{TMP01} for simulations in the
$\tau$CDM cosmology because there it was assumed that the emission is
purely bremsstrahlung, which underestimates cooling rates below $10^7$K
and places less weight on the innermost region of clusters.

In contrast to the non-radiative run, the clusters from the radiative
simulation have emission-weighted temperatures that show less scatter
and that exceed the virial values.  The reason for this is that the core
gas has cooled to low temperatures and been removed from the
intracluster medium, leaving behind higher-entropy, higher-temperature
gas---an effect that is more pronounced in lower-mass clusters.  The
emission is no longer dominated by the core and is well-resolved by
our simulations.  The clusters provide an adequate fit to the
observational data given the large uncertainty in the latter as
evidenced by the various broken lines in Fig.~\ref{fig:tm200}.

\subsection{Luminosity-temperature relations}
\label{sec:lt}

\vspace{\baselineskip}
\epsfxsize=6.5cm
\centerline{\rotatebox{270}{\epsfbox{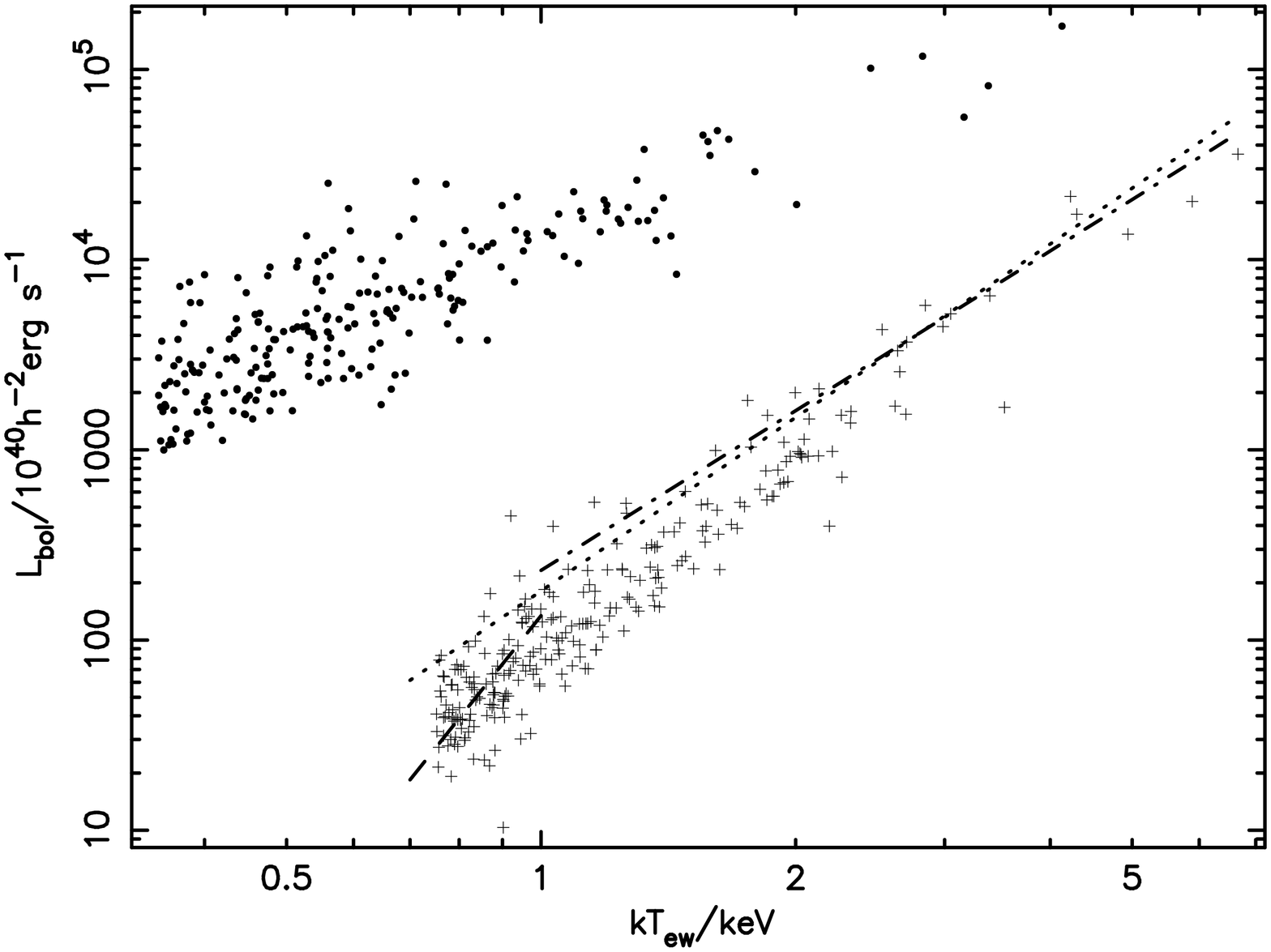}}}
\figcaption{The bolometric luminosity-temperature relations for our
simulations.  Clusters are represented by the same symbols used in
Fig.~\ref{fig:tm200} and broken lines represent the fits to
observational data by \citet{XuW00}.
\label{fig:lt200}}
\vspace{\baselineskip}

Fig.~\ref{fig:lt200} illustrates X-ray luminosity-temperature
relations for both simulations, again using crosses and filled circles
for radiative and non-radiative clusters respectively.  We have
trimmed the original catalogues by selecting clusters only with
temperatures above 0.35 and 0.75\,keV for the non-radiative and
radiative cooling catalogues respectively, to assure completeness in
temperature.  The broken lines illustrate best-fit power-law relations
as determined by \citet{XuW00} for their collected group sample
(dashed line), cluster sample (dot-dashed line) and combined sample
(dotted line).  (Note that the units for the normalisation of the
relations in Table~2 of Xue \& Wu 2000 are misquoted and should be a
factor of 10 larger, in agreement with their Table~1 and Figure~1.)

A power-law fit to results from the radiative simulation gives
\begin{equation}
L_{\rm bol} = 9.0\times10^{41}
\left(\frac{T_{\rm ew}}{1\,{\rm keV}}\right)^{3.3} 
                \, h^{-2} \, {\rm erg \, s^{-1}},
\label{eq:lxtxcool}
\end{equation}
and for the non-radiative simulation
\begin{equation}
L_{\rm bol} = 1.2\times10^{44}
\left(\frac{T_{\rm x}}{1\,{\rm keV}}\right)^{1.9} 
                \, h^{-2} \, {\rm erg \, s^{-1}}.
\label{eq:lxtxnocool}
\end{equation}

(As stated in section \ref{sec:mt}, these fits should not be taken as
precise calibrations.)
Again, both the slope and normalization differ between the two
simulations.  The non-radiative simulation gives a slope close to 2,
as predicted by self-similar scaling relations (the relation is
slightly flatter due to the inclusion of line emission). However, the
radiative simulation gives a significantly steeper slope of $3.3$.
The normalization of the relation is around a factor of 10--100 lower
in the radiative simulation than in the non-radiative simulation.
Qualitatively, these differences are the same as were found by
\citet{PTC00}, for a smaller cluster sample.  The high entropy gas
that replaces cooled material in the radiative simulation is hotter and less
dense than the gas in the non-radiative simulation.
The change in density has the greater effect, since the X-ray emissivity is
a slow function of temperature but is 
proportional to  the square of the gas density.   The combination
of the increase in temperature and the decrease in luminosity of the
clusters causes the substantial shift in the $L_{\rm bol}-T_{\rm ew}$
relation.

The results from the radiative simulation are in good agreement with
the best-fit relations of \citet{XuW00}.  If anything, we predict
luminosities that are are a factor of 2--3 too low, although there is still
significant uncertainty in the observational determinations.

\section{Conclusions}
\label{sec:conclusions}

In this letter, we have presented results from an ongoing programme to
measure the evolution of X-ray cluster properties for a range of
physical and cosmological models. Specifically, we have presented the
current-day ($z=0$) emission-weighted temperature-mass ($T_{\rm
ew}-M$) and bolometric luminosity-temperature ($L_{\rm bol}-T_{\rm
ew}$) relations from two simulations of a $\Lambda$CDM cosmology, one
with and one without radiative cooling.

The $T_{\rm ew}-M$ relation is significantly different in non-radiative
and radiative simulations, with the latter in reasonable agreement with
observational determinations.   In the non-cooling simulation, the
emission is dominated by cold, dense gas in the cores of the clusters;
radiative cooling removes this gas from the intracluster medium
(converting it into stars) and replaces it with higher-entropy, hotter
material.  This effect is more prevalent in lower-mass systems and so
flattens the temperature-mass relation.

The $L_{\rm bol}-T_{\rm ew}$ relation is also significantly different
between the two simulations. The high entropy material in the
radiative simulation is less dense than the material it replaces and
causes the X-ray luminosity of clusters to decrease by around a factor
of 100 at $T_{\rm ew} = 1$\,keV. The slope of the relation in the
non-radiative simulation is 1.9, similar to self-similar
predictions. However, the slope of the radiative relation is
significantly steeper, 3.3, again due to the differential effect of
cooling with halo temperature.  The radiative simulation is in much
closer agreement with the observations, both in the slope and
normalization of the relation.

In this paper, we have used bolometric luminosities and
emission-weighted temperatures.  For the low-temperature clusters
found in the non-radiative simulations, these will differ
significantly from properties measured in any particular X-ray band.
However, we wish to emphasize that to attempt to correct for this is
misleading as the use of such simulations is wrong in
principle---the low-entropy gas in the cores of these clusters has a
short cooling time and will not be present in real systems.

The global fraction of cooled gas (and stars) in the radiative simulation, 
20 per cent, is higher than suggested by
observations of the $K$-band galaxy luminosity function \citep{BPB01}. More 
importantly, the fraction of cooled gas within the clusters varies between 
about one-third and two-thirds with decreasing mass. While these values are 
not convincingly ruled out by observations, most people would also regard
these as high values---in which case our simulation can be treated as an
upper bound on the effect of radiative cooling.

In this paper, we have deliberately ignored the effect of
non-gravitational heating upon the gas.  In reality, we know that
there must be heating associated with star-formation and
metal-enrichment of the intracluster medium.  This will raise the
entropy of the gas and reduce the amount of cooling that is required
to match the observations. However, we do not regard the case
for significant heating by AGN or very efficient supernovae feedback as
proven. 

\acknowledgements
The simulations described in this paper were carried out on the
Cray-T3E at the Edinburgh Parallel Computing Centre as part of the
Virgo Consortium investigations of cosmological structure formation.
OM is supported by a DPST Scholarship from the Thai government; PAT is
a PPARC Lecturer Fellow.


\end{document}